\documentstyle[preprint,prb,aps]{revtex}
\begin{document}
\draft
\title{Small-$q$ Anomaly in the Dielectric Function and High-Temperature Oscillations of  Screening Potential in  2D Electron Gas with Spin-Orbit Coupling}

\author{Guang-Hong Chen and M.~E.~Raikh}
\address{Department of Physics, University of Utah, 
Salt Lake City, Utah  84112}


\maketitle

\begin{abstract}
\baselineskip=0.90cm
We study the static dielectric function $\varepsilon(q)$ of 2D electron system with spin-orbit coupling in the frame of the random phase approximation. We demonstrate that, in addition to the well-known $2k_F$-Kohn anomaly, spin-orbit coupling gives rise to the novel anomaly 
in the dielectric function at small $q=q_0\ll k_F$, where $q_0$ is the distance between two Fermi surfaces. 
As a result of this anomaly a large-distance behavior of the
potential from a point charge exhibits (in addition to the 
conventional Friedel oscillations) novel oscillations with a period
$2\pi/q_0$. 
The remarkable feature of these oscillations is that they are not smeared out by the temperature. We show that the
small-$q$ anomaly also modifies the indirect exchange interaction 
of localized magnetic moments (RKKY interaction). 
In the presense of spin-orbit coupling this interaction acquires a high-temperature component. 
\end{abstract}

\pacs{ PACS number(s): 71.45. Gm,73.20-r, 73.20.Dx}


\section{Introduction}
The dielectric function of a  two-dimentional electron system was first evaluated by F. Stern [1] within the random phase approximation (RPA). In the static case, Stern obtained the following expression for $\varepsilon(q)$
\begin{equation}
\label{stern}
\varepsilon(q)=\varepsilon_0(1-v(q)F(q)),
\end{equation}
where $q$ is the wave vector and $v(q)=2\pi e^2/\varepsilon_0q$ is the Fourier component of the Coulomb interaction, $\varepsilon_0$ is the dielectric constant of the medium. The response function $F(q)$ is given by
\begin{equation}
\label{f1}
F(q)=-\nu       \hspace{5.0cm}(q\leq2k_{F})
\end{equation}
\begin{equation}
\label{f2}
F(q)=-\frac{\nu}{q}(q-\sqrt{q^2-4k_F^2}) \hspace{2.0cm}(q>2k_{F}),
\end{equation}
where $\nu=m^*/\pi \hbar^{2}$ is the 2D density of states; $m^*$ is the effective  mass. It is well-kown that the singularity in $F(q)$ at $q=2k_{F}$ leads to the Friedel oscillations of the screening potential. For a point charge at a distance $d$ from a 2D plane the oscillatory part of the potential has the form

\begin{equation}
\label{osc}
V_{osc}(\rho)=-\biggl( \frac{2e^2}{\varepsilon_0a_{B}}\biggr) \frac{\sin(2k_{F}\rho)}{(2k_{F}\rho)^2}e^{-2k_Fd} ,
\end{equation}
where $a_{B}=\hbar^2\varepsilon_0/m^*e^2$ is the effective Bohr radius. Eq. (\ref{osc}) applies at large distances, $k_F\rho\gg 1$. The result (\ref{osc}) is derived for zero temperature $T=0$.
Obviously,  the oscillations are smeared out as $T$ increases. This effect amounts to an additional factor $A(u)$ in Eq. (\ref{osc}) defined as
\begin{equation}
\label{a}
A(u)=\frac{u}{\sinh(u)},
\end{equation}
where $u=\pi k_{F} \rho T/E_{F}$. It follows from Eq. (5) that the power-law decay of $V_{osc}(\rho)$ persists at finite temperature from $\rho\sim k_{F}^{-1}$ to $\rho\sim k_{F}^{-1}(E_{F}/T)$. At larger $\rho$ a crossover to the exponential decay $\propto exp(-2\pi k_{F} \rho
 T/E_{F})$ takes place.

The expression for $V_{osc}(\rho)$  can be straightforwardly  generalized to the case of two or more 2D subbands. This is because the size-quantization wave functions in the direction perpendicular to the plane are orthogonal, so that the inter-subband transitions are forbidden. In this case $V_{osc}(\rho)$ is described by  the sum of the Friedel oscillations from  each subband. 

The same argument would apply in the presence of the Zeeman splitting if one neglects the orbital action of the magnetic field. In this case, the transitions between the  branches of the spectrum with different spin projections are forbidden.

In the present paper we demonstrate that the situation is quite different in the presence of  the spin-orbit (SO) coupling. With SO coupling, the electronic states are classified not by the spin orientation but by the {\em chirality}. Our main results are as follows: (i) SO coupling gives rise to a new type of oscillations in $V(\rho)$ which are caused by virtual transitions between the states with different chirality; (ii) the distinguishing feature of these oscillations is that they survive the increasing of  temperature (not smeared out with the increasing $T$).

To account for SO coupling, we choose the simplest form of the SO interaction for 2D electron system [2]
\begin{equation}
\label{hso}
\hat{H}_{SO}=\alpha \bf{k}\cdot( \bbox{\sigma} \times \bf{n}),
\end{equation}
where $\bf{k}$ is the momentum, $\bf{n}$ is the unit vector normal to the 2D plane, $\bbox{\sigma}=(\sigma_{1},\sigma_{2},\sigma_{3})$ are the Pauli matrices, and $\alpha$ is the SO coupling constant.
With spin-orbit term Eq. (6) the splitted Fermi surfaces for a {\em given} 2D subband represents two circles with radii (see Fig. 1)
\begin{equation}
\label{kf1}
k_{F}^{(1)}=-\frac{q_0}{2}+\sqrt{\frac{q_0^2}{4}+\frac{2m^*E_{F}}{\hbar^2}}=k_F-\frac{q_0}{2}
\end{equation}
\begin{equation}
\label{kf2}
k_{F}^{(2)}=\frac{q_0}{2}+\sqrt{\frac{q_0^2}{4}+\frac{2m^*E_{F}}{\hbar^2}}=k_F+\frac{q_0}{2},
\end{equation}
where $k_F$ is related to the Fermi energy as
\begin{equation}
\label{kf}
k_F=\sqrt{\frac{q^2_0}{4}+\frac{2m^*E_F}{\hbar^2}},
\end{equation}
and $q_0=k_{F}^{(2)}-k_{F}^{(1)}$ is expressed through the SO coupling constant as follows
\begin{equation}
\label{q0}
q_0=\frac{2m^*\alpha}{\hbar^2}.
\end{equation}
It is important that the difference $k_F^{(2)}-k_F^{(1)}$ does not depend on the position of the Fermi level.

Our main observation is that, in contrast to the case of {\em different} 2D subbands, the virtual transitions between SO-splitted states within the {\em same} subband are allowed and, thus, {\em do} {\em  contribute} to the response function. Below we will demonstrate that this contribution is singular at $q=q_0$. This singularity results in the following oscillatory behavior of the screening potential
\begin{equation}
\label{vso}
V_{SO}(\rho)=-\biggl(\frac{2e^2}{\varepsilon_0a_{B}}\biggr) \biggl(\frac{q_0}{2k_{F}}\biggr)^2\frac{\cos(q_0\rho)}{(q_0\rho)^2} .
\end{equation}
Comparing (\ref{vso}) to  (\ref{osc}), one can see that the SO-induced oscillations do not contain the factor $exp(-2k_Fd)$. Strictly speaking, this factor should be replaced by $exp(-q_0d)$ in Eq. (\ref{vso}), but we assume throughout this paper that SO coupling is weak, i.e.  $q_0\ll k_F$. The  magnitudes of the oscillations Eq. (\ref{vso}) and Eq. (\ref{osc}) are of the same order  for $d=0$. However, at elevated temperatures, $V_{SO}$ will always dominate since the conventional Friedel oscillations get damped.

It is interesting to note that as far as  the conventional Friedel oscillations are concerned,  there is also a difference between the case of two size-quantization and two SO-splitted subbands. In the former case the large-$\rho$ behavior of the screened potential is $V(\rho)\propto (\sin(2k_F^{(1)}\rho)+\sin(2k_F^{(2)}\rho))/\rho^2$, whereas in the latter case, as we will demonstrate below,  these oscillations have the form $V(\rho)\propto (\sin(k_F^{(1)}+k_F^{(2)})\rho)/\rho^2 $.

The paper is  organized as follows: In Sec. II we evaluate the dielectric function $\varepsilon(q)$ for different domains of wave vector $q$.  The spatial oscillations of the screened potential are studied  in Sec. III. Concluding remarks are given in the Sec. IV.

\section{Dielectric Function with SO coupling}

\subsection{Spectrum and Eigenfunctions}

With  SO term Eq. (\ref{hso}), the Hamiltonian of a free electron has the form

\begin{equation}
\label{hamiltonian}
H=\frac{\hbar^2k^2}{2m}+\alpha \bf{k}\cdot(\bbox{\sigma}\times\bf{n}).
\end{equation}

The energy spectrum and the wave functions are given by

\begin{equation}
\label{spectrum}
E_{\mu}(k)=\frac{\hbar^2k^2}{2m}-\mu\alpha k
\end{equation}
\begin{eqnarray}
\label{spinor}
\chi_{{\bf k}\mu}=\frac{1}{\sqrt{2}}\left[\begin{array}{cc}i\mu e^{-i\phi_{\bf{k}}}\\1\end{array}\right] , \hspace{3.0cm}
\Psi_{{\bf k},\mu}(\bbox{\rho})= \frac{1}{\sqrt{A}}
  e^{i {\bf k}\cdot{ \bbox{\rho}}} \chi_{{\bf k}\mu}.
\end{eqnarray}

Here $\mu=\pm 1 $ is the  chirality,  $A$ is the normalization area, and $\phi_{\bf{k}}$ is the azimuthal angle of the vector $\bf{k}$.

\subsection{General Formula for the Dielectric Function}

We start with the RPA formula for $\varepsilon(q)$ applied to the case when free electron wave functions are the eigenfuctions of the Hamiltonian Eq. (\ref{hamiltonian})

\begin{equation}
\label{die}
\frac{\varepsilon(q)}{\varepsilon_0}=1-v(q)\sum_{\bf{k},\mu,\mu^{'}}|\chi^{\dagger}_{\bf{k},\mu}\chi_{\bf{k+q},\mu^{'}}|^2 \frac{n(E_{\mu}(k))-n(E_{\mu^{'}}(|\bf{k}+\bf{q}|))}{\em E_{\mu}(k)-E_{\mu^{'}}(|\bf{k}+\bf{q}|)},
\end{equation}
where $n(E_{\mu}(k))$ is the Fermi-Dirac distribution function, and 
\begin{equation}
\label{overlap}
|\chi^{\dagger}_{\bf{k},\mu}\chi_{\bf{k+q},\mu^{'}}|^2=\frac{1}{2}(1+\mu\mu^{'}\cos(\phi_{\bf{k}+\bf{q}}-\phi_{\bf{k}})),
\end{equation}
are the overlap integrals. It is important to note that the virtual transitions between different chiralities are {\em allowed}.
 In the case of two size-quantization subbands, the overlap integral would {\em reduce to the Kronecker symbol}, so that $\varepsilon(q)$ would represent a sum of contributions from each subband. In contrast, as it follows from Eq. (\ref{overlap}), the overlap integral vanishes for $\mu\neq\mu^{'}$ {\em only} if $\bf{k}\parallel\bf{q}$.

In evaluating the dielectric function, it is convenient first to perform the angular integration  by introducing the auxilary variable in Eq. (\ref{die})
\begin{eqnarray}
\label{aux}
\frac{\varepsilon(q)}{\varepsilon_0}&=& 1-\frac{v(q)}{4\pi^2}\sum_{\mu,\mu^{'}}\int_0^{\infty} dk k \int_0^{\infty} dp p\int_0^{2\pi} d \phi_{\bf{k}} \biggl(1+\mu\mu^{'}\frac{\bf{k}\cdot(\bf{k}+\bf{q})}{\em kp}\biggr)\times \nonumber \\
              & & \biggl(\frac{n(E_{\mu}(k))-n(E_{\mu^{'}}(p))}{E_{\mu}(k)-E_{\mu^{'}}(p)}\biggr) \delta(p^2-|{\bf k}+{\bf q}|^2)\nonumber \\
              &=& 1-v(q)\sum_{\mu,\mu^{'}}F_{\mu\mu^{'}}(q),
\end{eqnarray}
where we introduced the following notation
\begin{eqnarray}
\label{fuu1}
F_{\mu\mu^{'}}(q)&=&\frac{1}{4\pi^2}\int dk k \int dp p\int d \phi_{\bf{k}} \biggl[1+\mu\mu^{'}\frac{\bf{k}\cdot\bf{(k+q)}}{\em kp}\biggr]\times \nonumber \\
                 & &\biggl[\frac{n(E_{\mu}(k))-n(E_{\mu^{'}}(p))}{E_{\mu}(k)-E_{\mu^{'}}(p)}\biggr] \delta(p^2-|{\bf k}+{\bf q}|^ 2).
\end{eqnarray}
Integrating over $\phi_{\bf{k}}$, we obtain
\begin{equation}
\label{fuu2}
F_{\mu\mu^{'}}(q)=\frac{1}{4\pi^2}\int dk \int dp \frac{n(E_{\mu}(k))-n(E_{\mu^{'}}(p))}{E_{\mu}(k)-E_{\mu^{'}}(p)}\biggl(\frac{q^2-(k-p)^2}{(k+p)^2-q^2}\biggr)^{-\frac{\mu\mu^{'}}{2}}.
\end{equation}
It immediately follows from symmetry that $F_{1,-1}=F_{-1,1}$. The further caculations are different for the cases $\mu=\mu^{'}$ and $\mu=-\mu^{'}$.  In the next subsections we consider these cases seperately.

\subsection{ The case  $\mu=-\mu^{'}$}
 The expression for $F_{1,-1}(q)$ can be conveniently rewritten in the form
\begin{equation}
\label{f1-1}
F_{1,-1}(q)=\frac{\nu}{2\pi}\int_0^{\infty}dk[n(E_{+1}(k))G_+(k,q)+n(E_{-1}(k))G_-(k,q)],
\end{equation}
where the function $G_{\pm}(k,q)$ is defined as
\begin{equation}
\label{gpm}
G_{\pm}(k,q)=\int_0^{\infty}\frac{dp}{(k+p)(k-p\pm q_0)}\sqrt{\frac{q^2-(k-p)^2}{(k+p)^2-q^2}}.
\end{equation}
It is presumed that the integrand is zero when the expression under the square root is negative.

Consider now the case of small $q\ll k_F$.  Then the actual limits of integration in Eq. (\ref{gpm}) are $(k-q, k+q)$, i.e. $p$ is close to $k$. This allows to evaluate Eq. (\ref{gpm}) analytically
\begin{equation}
\label{gkq}
G_{\pm}(k,q)\approx \frac{1}{4k^2}\int_{k-q}^{k+q} dp \frac{\sqrt{q^2-(k-p)^2}}{k-p\pm q_0}=\pm\frac{\pi}{4k^2}\biggl[q_0-\sqrt{q_0^2-q^2} \theta(q_0-q)\biggr],
\end{equation}
where $\theta(x)$ is a step-function.

We see that in the small-$q$ limit, the $q$ and $k$ dependencies in the $G(k,q)$  get decoupled from each other so that the $q$-dependence of the response function $F_{1,-1}(q)$ is the same as the $q$-dependence of Eq. (\ref{gkq}):
\begin{equation}
\label{f1-1q}
F_{1,-1}(q)=\frac{\nu}{8}\biggl[q_0-\sqrt{q_0^2-q^2}\theta(q_0-q)\biggr]\int_0^{\infty} dk \frac{n(E_{+1}(k))-n(E_{-1}(k))}{k^2}
\end{equation}
First, we realize  that the response function exhibits a square root singularity at $q=q_0$. This singularity is depicted schematically in Fig. 2a. The second important observation is that,  since $q_0$ is determined by the parameters of the electron spectrum and does not depend on the Fermi level position (see Eq. (\ref{q0})), the temperature dependence of $F_{1,-1}(q)$ is very weak. Indeed, the straightforward calculation of the integral over $k$ yields
\begin{equation}
\label{fqt}
F_{1,-1}(q,T)=-\frac{\nu q_0}{8k_F^2}\biggl[q_0-\sqrt{q_0^2-q^2}\theta(q_0-q)\biggr]\biggl[1+\frac{\pi^2}{6}\biggl(\frac{T}{E_F}\biggr)^2\biggr].
\end{equation}

Such temperature dependence should be contrasted to that  of the conventional $2k_F$-anomaly. At  finite $T$, the anomalous term,  $\sqrt{q^2-4k_F^2}$, in Eq. (\ref{f2}) is replaced by [3,4]
\begin{equation}
\frac{1}{2}\int  dx \frac{\sqrt{q^2-4k_F^2+8m^*Tx/\hbar^2}}{\cosh^{2}x},
\end{equation}
which leads to the smearing of the anomaly within the region $|q-2k_F|\sim k_F T/E_F$. 

Let us  consider the case of large $q\sim \biggl(k_F^{(1)}+k_F^{(2)}\biggr)=2k_F$. In this case the limits of the integration in Eq. (\ref{gpm}) are $(q-k,\infty)$. By noting that the main contribution to the anomaly comes from the vicinity of $q\sim (k+p)$, the integral  can be simplified as
\begin{equation}
\label{appl}
G_{\pm}(k,q)\approx \frac{1}{\sqrt{2q}}\int_{q-k}^{\infty}\frac{dp}{(k-p\pm q_0)\sqrt{k+p-q}}=-\frac{\pi}{\sqrt{2q}}\frac{\theta(q-2k\mp q_0)}{\sqrt{q-2k\mp q_0}},
\end{equation}
Substituting Eq. (\ref{appl}) into Eq. (\ref{f1-1}) leads to the following results for the response function
\begin{equation}
\label{qsmall}
F_{1,-1}(q)=-\frac{\nu}{\sqrt{2}},\hspace{0.5cm} {\mbox for} \hspace{0.3cm}q<2k_F
\end{equation}
\begin{equation}
\label{qbig}
F_{1,-1}(q)=-\frac{\nu}{\sqrt{2}}\biggl(1-\sqrt{\frac{q}{2k_F}-1}\biggr), \hspace{0.5cm} {\mbox for} \hspace{0.3cm} q\ge  2k_F.
\end{equation}
Eqs. (\ref{qsmall}) and (\ref{qbig}) apply in the domain  $q-2k_F\ll k_F$. We conclude that the anomaly occurs at  $q=k_F^{(1)}+k_F^{(2)}=2k_F$. This should be contrasted to the anomalies coming from the terms with $\mu=\mu^{'}$ in Eq. (\ref{die}) (intrasubband transitions).
These anomalies are studied in the next subsection.

\subsection{The case  $\mu=\mu^{'}$}
 
In this case the anomalies occur at $q= 2k_F^{(1)}$ in $F_{1,1}(q)$ and at
$q= 2k_F^{(2)}$ in $F_{-1,-1}(q)$. To extract the  non-analytic behavior, we note that the main contribution to the integral Eq. (\ref{fuu2}) comes from 
 $k,p\sim k_{F} $ and $q\sim k+p$, so that we can simplify the
   expression for  $F_{1,1}(q)$ and $F_{-1,-1}(q)$ in the  following way
\begin{equation}
F_{1,1}(q)=\frac{\sqrt{2}\nu}{\pi\sqrt{q}(q+ q_0)}\int\int dk dp\frac{n(E_{+1}(k))\sqrt{k+p-q}}{k-p},
\end{equation}
\begin{equation}
F_{-1,-1}(q)=\frac{\sqrt{2}\nu}{\pi\sqrt{q}(q-q_0)}\int\int dk dp\frac{n(E_{-1}(k))\sqrt{k+p-q}}{k-p}.
\end{equation}
Then by straightforward calculation we get
\begin{equation}
\label{3/21}
F_{1,1}(q)=F_{-1,-1}(q)=\frac{\sqrt{2}\nu}{3} \hspace{0.5cm} {\mbox for} \hspace{0.3cm} q<2k_F^{(1,2)},
\end{equation}

\begin{equation}
\label{3/22}
F_{1,1}(q)=\frac{\sqrt{2}\nu}{3}\biggl[1-\biggl(\frac{q-2k^{(1)}_F}{2k^{(1)}_F}\biggr)^{3/2}\biggr], \hspace{0.5cm} {\mbox for}  \hspace{0.3cm} q\ge 2k_F^{(1)},
\end{equation}
\begin{equation}
\label{3/23}
F_{-1,-1}(q)=\frac{\sqrt{2}\nu}{3}\biggl[1-\biggl(\frac{q-2k^{(2)}_F}{2k^{(2)}_F}\biggr)^{3/2}\biggr]  \hspace{0.5cm}{\mbox for}  \hspace{0.3cm} q\ge 2k_F^{(2)}.
\end{equation}
We see that instead of conventional $\sqrt{q-2k_F}$ anomaly, we obtained much weaker anomalies $\propto\biggl(q-2k^{(1,2)}_F\biggr)^{3/2}$. The origin  of such a weakening is that for virtual transitions, $k^{(1,2)}_F-2q\rightarrow -k^{(1,2)}_F$,  which give rise to the conventional Kohn anomaly, the corresponding overlap integral, Eq. (\ref{overlap}), is  identically  zero in the presence of the SO coupling.

\section{Oscillations of the Screening Potential}

Consider a point charge $e$ at distance $d$ from the 2D plane. The potential created by this charge within the plane is given by [3]
\begin{equation}
\label{vrho}
V(\rho)=e\int_0^{\infty} dq\frac{e^{-qd}J_0(q\rho)}{\varepsilon(q)}.
\end{equation}
As we demonstrated in the previous section, the dielectric  function $\epsilon(q)$ exhibits anomalous behavior at small $q=q_0$ and also at $q=k^{(1)}_F+k^{(2)}_F$. Besides, there are weak anomalies at $q=2k^{(1)}_F$, $q=2k^{(2)}_F$ caused by the intrasubband virtual transitions.

The general approach for extracting the oscillating behavior of $V(\rho)$ is as follows. Suppose that $\epsilon(q)$ has an anomally at some $q=q_c$.  Then the integrand in Eq. (\ref{vrho}) should be expanded with respect to $\delta {\cal F}$ defined as 
\begin{equation}
\label{deltaf}
\delta{\cal F}(\kappa)=F(q_c+\kappa)-F(q_c),
\end{equation}
and the Bessel function should be replaced by its large-$\rho$ asymptotics. This leads to the following general formula for the oscillating part of $V(\rho)$
\begin{equation}
\label{general}
V(\rho)=\frac{e v(q_c)}{\varepsilon_0}\sqrt{\frac{2}{\pi q_c\rho}}\biggl[\cos(q_c\rho-\frac{\pi}{4})I_1-\sin(q_c\rho-\frac{\pi}{4})I_2\biggr]\exp(-q_cd),
\end{equation}
where the integrals $I_1, I_2$ are defined as
\begin{equation}
\label{i1}
I_1=\int d\kappa\delta{\cal F}(\kappa)\cos(\kappa\rho),
\end{equation}
\begin{equation}
\label{i2}
I_2=\int d\kappa\delta {\cal F}(\kappa)\sin(\kappa\rho).
\end{equation}

 By using the results obtained  in the previous sections, we can now easily carry out the calculations (Eq. (\ref{general})- Eq. (\ref{i2})).

(i) For anomaly at $q_c=q_0$, we have
\begin{equation}
\label{fq0}
\delta {\cal F}(\kappa)\approx \frac{\sqrt{2}\nu}{8}\frac{q_0^{3/2}}{k_F^2}\sqrt{-\kappa}
\end{equation}
and $\epsilon(q_c)\approx \epsilon_0$. This leads us to the spin-orbit oscillations Eq. (\ref{vso}).

(ii) For anomaly at $q_c=k^{(1)}_F+k^{(2)}_F=2k_F$, the form of $\delta {\cal F}$ was shown to be
\begin{equation}
\label{fq2kf}
\delta {\cal F}(\kappa)=\nu\sqrt{\frac{\kappa}{k_F}},
\end{equation}
where we have considered the contribution of the $F_{-1,1}(q)$ which gives the same contribution as $F_{1,-1}(q)$. It is worth  
noting that this form coincides with the standard form of the Kohn anomaly in 2D. Thus for the oscillations with period $\pi/k_F$ we get the same form of the screening potential as Eq. (\ref{osc}).

(iii) At  $q_c=2k^{(1,2)}_F$, it follows from Eqs. (\ref{3/22}) and (\ref{3/23}) that
\begin{equation}
\delta{\cal F}(\kappa)=\frac{\nu}{3}\biggl(\frac{\kappa}{2k^{(1,2)}_F}\biggr)^{3/2}.
\end{equation}
Substituting this form into Eqs. (\ref{general}-\ref{i2})  results in 
\begin{equation}
V_{osc}(\rho)=\frac{3\sqrt{2}e^2}{2\varepsilon_0a_B}\frac{\sin(2k^{(1,2)}_F\rho)}{(2k^{(1,2)}_F\rho)^3}e^{-2 k^{(1,2)}_F d}.
\end{equation}

We see that oscillations with ``usual'' periods $\pi/k^{(1,2)}_F$ fall off as $1/\rho^3$, i.e. faster than the conventional oscillations Eq. (\ref{osc}).

\section{Conclusion}

The main result of the present paper is the prediction of the novel oscillations of the screening potential originating from the spin-orbit term in the Hamiltonian of a free electron.  In our calculations we
assumed that the bare energy spectrum is parabolic. We also assumed that the origin of the spin-orbit term is the asymmetry of the confinement potential [2] and neglected the  term coming from the absence of the inversion symmetry in the bulk [5]. 
Note, however, that if the Dresselhaus mechanism [5] dominates the
SO coupling (in this case $\hat H_{SO}\propto (\sigma_{x}k_{x}-\sigma_{y}k_{y})$), the results of the present paper remain unchanged.

 There are two cases when the spin-orbit-induced
oscillations,  studied in the present paper,  dominate over the conventional $2k_F$-oscillations [1].
First,  with increasing temperature, $2k_F$-oscillations get damped,
 while $q_0$-oscillations remain almost temperature-independent. This situation is illustrated in Fig. 3. It is seen that for parameters of Fig. 3 the Friedel oscillations almost disappear at $T\approx 0.1E_F$.
Second,  with increasing the spacing $d$ from the ionized impurity to the 2D electron plane the amplitude of the conventional oscillations falls off as $exp(-2k_Fd)$ [1], whereas for $q_0$-oscillations this dependence is much weaker   $\propto exp(-q_0d)$. Fig. 2 illustrates how the $q_0$-oscillations take over as $d$ is increased. Note,  that large values of the spacing $d$ is a necessary requirement for  high mobility of 2D electron systems [6]. On the other hand, we considered the ideal electron gas, in the sense that  we assumed  the condition $q_0l\gg 1$, $l$ being the mean free path, to be met. Thus, the  large values $d$ is the
domain of validity of our results.

Another  application of the results obtained  in the
present paper  is the effect of the SO coupling on indirect
exchange interaction of localized magnetic moments (RKKY interaction).
It is straightforward to generalize the expression for RKKY in two
dimensions [7]  to  include the  SO coupling
\begin{equation}
\label{rkky}
H_{RKKY}(\rho_{12})=J^2\biggl[\Phi_1(\rho_{12}){\bf S}_1\cdot{\bf S}_2+(\Phi_2(\rho_{12})-\Phi_1(\rho_{12}))({\bf S_1}\cdot{\bf n})({\bf S}_2\cdot{\bf n})\biggr], 
\end{equation}
where $\rho_{12}$ is the distance between the two localized spins ${\bf S}_1$
and ${\bf S}_2$, $J$ is the contact interaction constant, and the functions $\Phi_1(\rho)$, $\Phi_2(\rho)$ are defined as
\begin{equation}
\label{a12}
\Phi_1(\rho)=\sum_{\mu\mu^{'}}\int\frac{d^2q}{(2\pi)^2}e^{i{\bf q}\cdot\rho}\Pi_{\mu\mu^{'}}(q),
\end{equation}
\begin{equation}
\label{b12}
\Phi_2(\rho)=\sum_{\mu\mu^{'}}\int\frac{d^2q}{(2\pi)^2}e^{i{\bf q}\cdot\rho}F_{\mu\mu^{'}}(q).
\end{equation}
In Eq. (\ref{b12})  the response function $F_{\mu\mu^{'}}(q)$
 is defined by Eq. (\ref{fuu1}), whereas  $\Pi_{\mu\mu^{'}}(q)$ is the conventional polarizability  
\begin{equation}
\label{puu}
\Pi_{\mu\mu^{'}}(q)=\sum_{\bf k}\frac{n(E_{\mu}(k))-n(E_{\mu^{'}}(|\bf{k}+\bf{q}|))}{\em E_{\mu}(k)-E_{\mu^{'}}(|\bf{k}+\bf{q}|)}.
\end{equation}
In the absence of the SO coupling we have $\Phi_1(\rho)=\Phi_2(\rho)$, and the
interaction falls off at large distances as $\sin(2k_F\rho)/\rho^2$.
As it was demonstrated above,  with SO coupling the function $\Phi_1(\rho)$ contains high-temperature oscillations, $\cos(q_0\rho)/\rho^2$ 
 resulting from the term $F_{1,-1}(q)=F_{-1,1}(q)$ in the r.h.s  of Eq. (\ref{a12}). The function $\Phi_2(\rho)$ also exhibits high-temperature oscillations due to the term $\Pi_{1,-1}(q)=\Pi_{-1,1}(q)$ in the r.h.s of Eq. (\ref{b12}). The small-$q$ anomaly in this term can be extracted in the similar way as it was done in Sec. II. We list only the final
result
\begin{equation}
\label{p}
\Pi_{1,-1}(q)= \frac{\nu \theta (q_0-q)}{4\sqrt{q_0^2-q^2}}.
\end{equation}
It is seen that the anomaly at $q=q_0$ in $\Pi_{1,-1}$ is
much stronger than in $F_{1,-1}$. As a result, at large distances,   $\Phi_2(\rho)$ is much bigger (by a factor $\sim k_F^2\rho/q_0$) than
$\Phi_1(\rho)$. Then the $\rho$-dependence of  the
indirect exchange  interaction takes the form $\sin(q_0\rho)/\rho$.
As it follows from Eq. (\ref{rkky}) the in-plane components of 
${\bf S}_1$ and ${\bf S}_2$ interact at large $\rho$ much stronger
than the normal components.

As a final remark, we address the above assumption that the SO splitting
of the Fermi surface does not depend on $k_F$. Strictly speaking, since 
$k_F$ is determined by the electron concentration $n$, and, in turn,
the change of $n$ results in the change of the confinement potential,
there should be a certain dependence of $\alpha$ on $n$.  Recent 
magnetotransport experiments on different 2D electronic systems
yield either $\alpha$ weakly changing with $n$ (Refs. [8], [9]) or, 
$\alpha$,
roughly, inversely proportional to $n$ (Ref. [10]). Meanwhile, it
is important to note that $\alpha$ depends only on the {\em total}
concentration of electrons, which does not depend on temperature.
Thus, our main conclusion that spin-orbit-induced anomaly
in $\varepsilon(q)$ is not smeared out by temperature is 
insensitive to the actual dependence $\alpha(n)$. This dependence can govern only the position
of the anomaly. 
\vspace{0.5cm}
\begin{flushleft} \Large \bf
Acknowledgements
\end{flushleft}
The authors are strongly grateful to  T. V. Shahbazyan for careful reading  the manuscript and many valuable suggestions. We  acknowledge the stimulating  discussions with ~ R. R. Du, E. I. Rashba, and Y. S. Wu. 
Discussions with H. U. Baranger and Y. Imry are also gratefully acknowledged.
One of the authors (G.H.C.) thanks  A. Galstyan for technical
 assistance in preparation of the manuscript. M.E.R. is grateful to
the Aspen Center for Physics for hospitality.

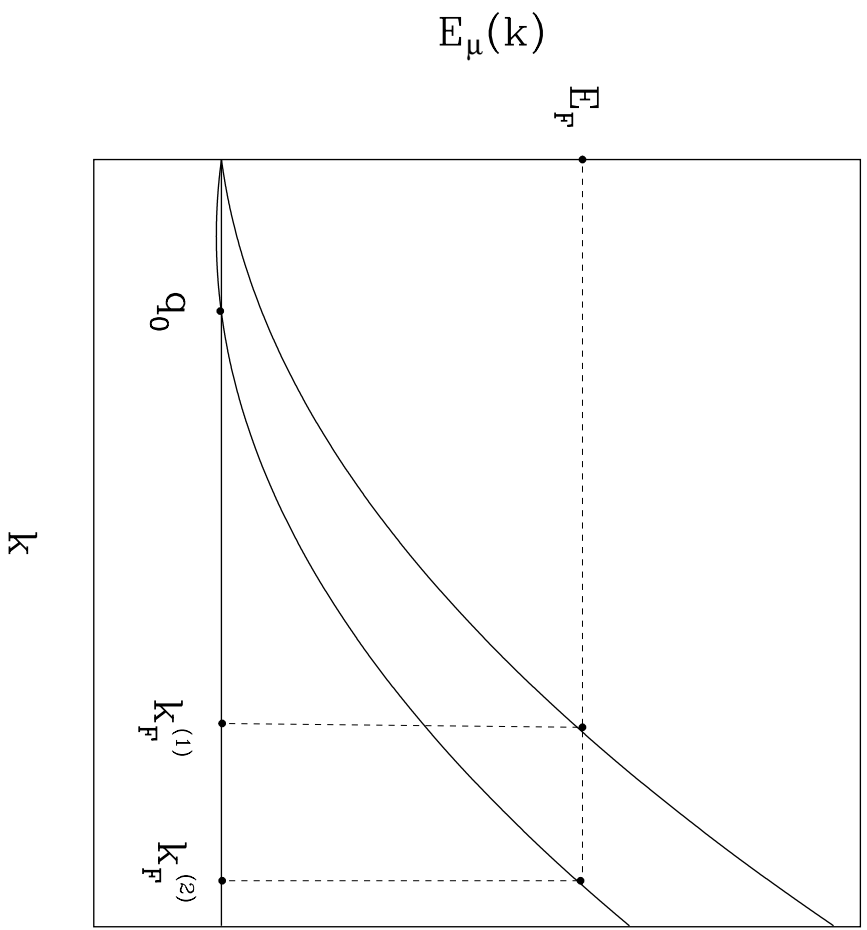
\begin{figure}
\caption{ The energy spectrum of a 2D system with spin-orbit coupling;
upper and lower branches correspond to  chiralities $\mu =-1$ and
$\mu=1$ respectively;
 the difference between two Fermi momenta $k_F^{(2)}$ and $k_F^{(1)}$ is $q_0$ determined by Eq. (10). }
\end {figure}
\begin{figure}
\caption{The numerical results for large-$\rho$ behavior of dimensionless screened potential $V(\rho)(\frac{4e^2}{\varepsilon_0a_B})^{-1}$ at $T=0$ are plotted for different distances, $d$, of a point charge from the 2D plane at $q_0/2k_F=1/4$. (a) $d=a_B$. The solid curve represents  the result calculated within the  
Thomas-Fermi approximation:
$\varepsilon (q)=\varepsilon_0(1+2/qa_B)$.
Dotted  curve is the RPA result in the absence of the SO coupling featuring
the Friedel oscillations. Dashed-dotted curve is plotted with $F(q)$
in the form of Eq. (23) and features the $q_0$-oscillations.
 The inset shows schematically the small-$q$ anomaly in the response function $F(q)$. (b) the same as (a) for $d=1.5a_B$. (c) the oscillating parts of the screened potential for $d=a_B$ and $d=1.5a_B$. Dotted and dashed-double dotted curves correspond to the Friedel oscillations. Thick solid and thick dashed-dotted curves correspond to the $q_0$-oscillations. }
\end{figure}
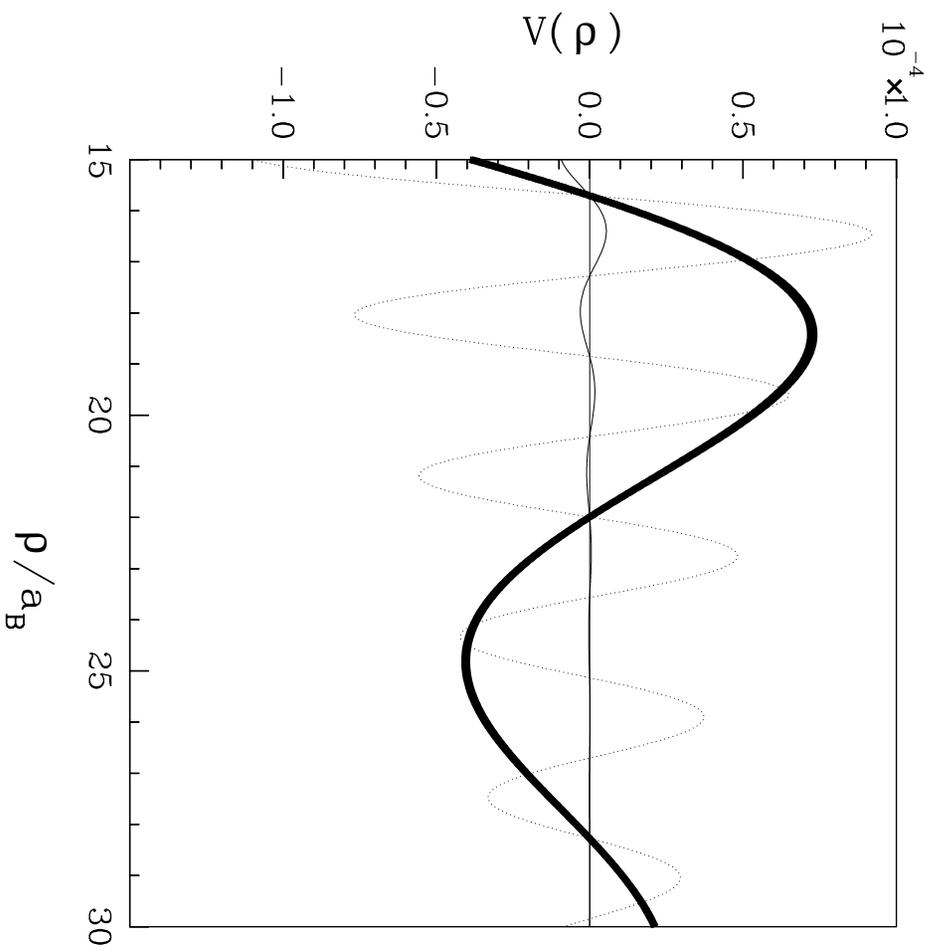
\begin{figure}
\caption{ Asymptotics of the screened  potential $V(\rho)$ in the units 
of $2e^2/\varepsilon_0a_B$
 are  plotted at different temperatures at $q_0/2k_F=1/4$. Dotted and solid curves show the $2k_F$-oscillations at  $T=0$ and  $T=0.1E_F$ respectively. Thick curve represents the  $q_0$-oscillations, Eq. (11),  their change with temperature from $T=0$ to $T=0.1E_F$ being smaller than the thickness of the curve.}
\end{figure}

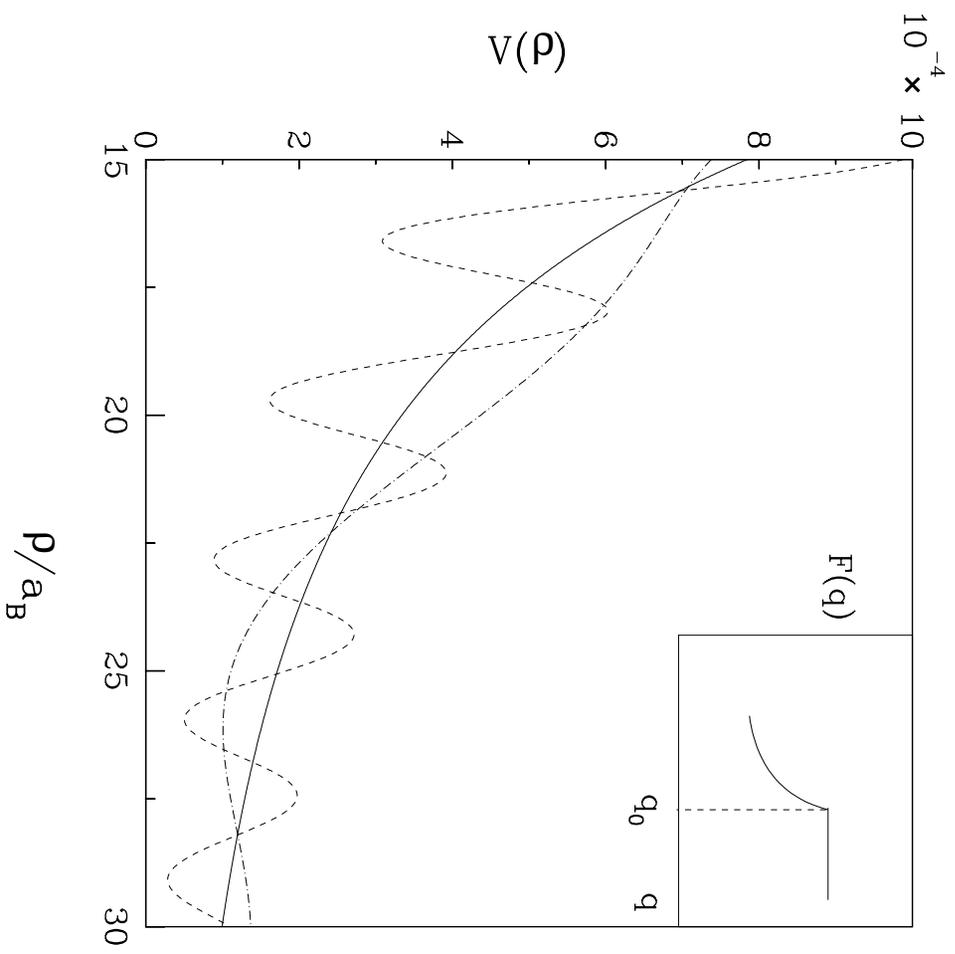
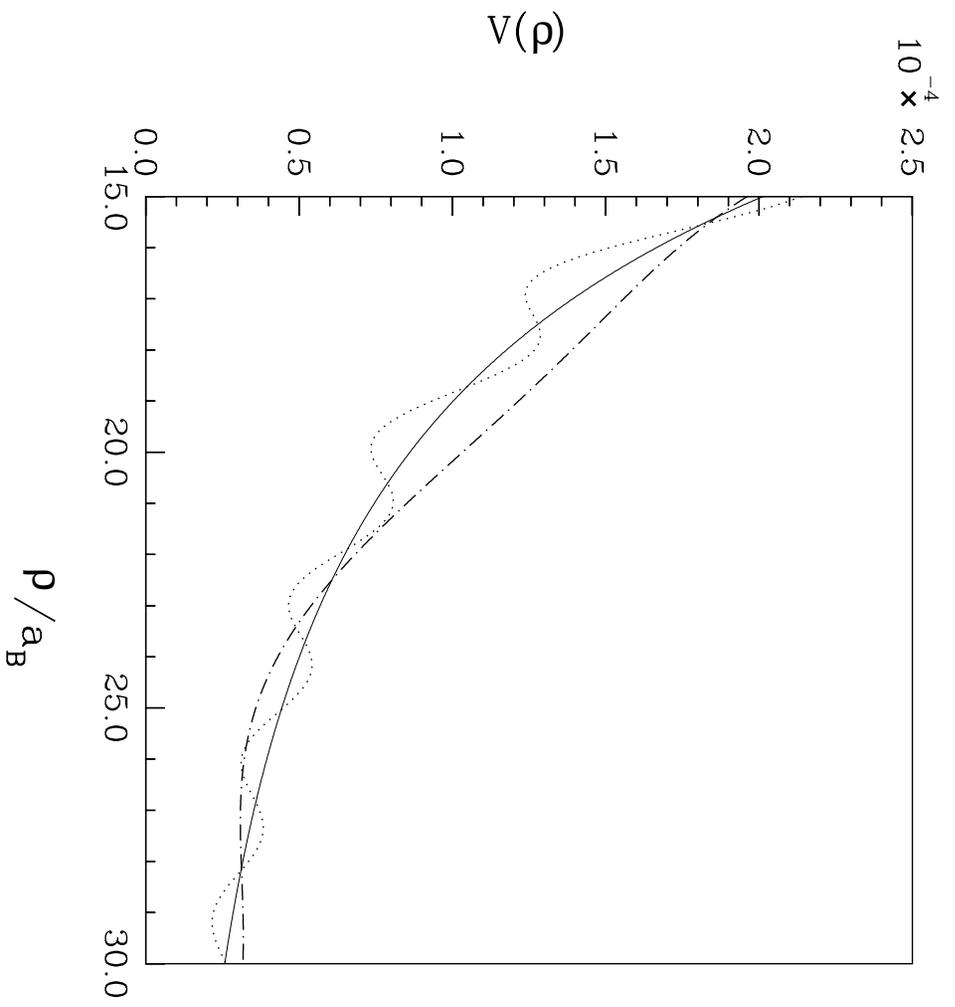
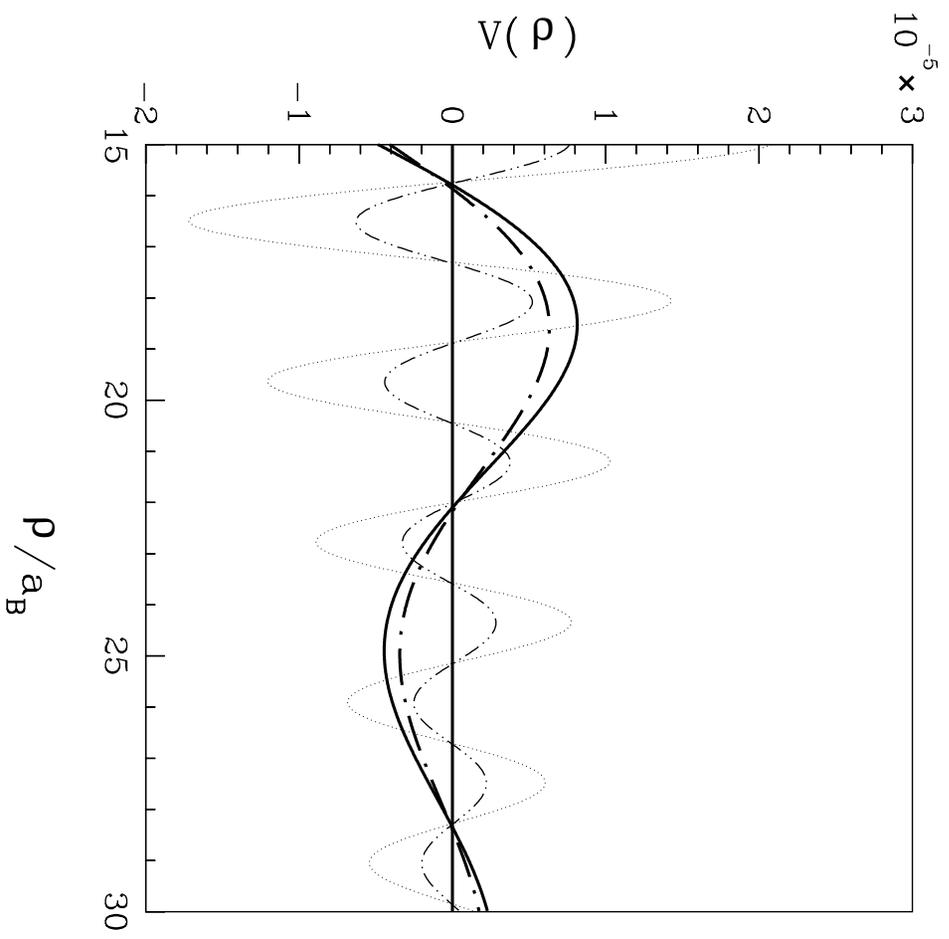
\end{document}